\documentclass[a4paper]{article}

\usepackage[dvips]{graphicx}
\usepackage{pst-all,url}
\usepackage{amsmath,amssymb}
\usepackage[thmmarks]{ntheorem}

\newcommand{\Rel}{\mathbb{Z}}
\newcommand{\Nat}{\mathbb{N}}
\renewcommand{\mod}{\textrm{ \small mod }}

\newrgbcolor{notwhite}{0.9 0.9 0.9}
\newrgbcolor{notgray}{0.63 0.63 0.63}
\newrgbcolor{lightred}{1 0.6 0.6}
\newrgbcolor{lightblue}{0.4 0.4 1}
\newrgbcolor{lightgreen}{0.3 0.7 0.7}
\newrgbcolor{gold}{0.9 0.7 0.1}

\theoremstyle{break}
\theoremheaderfont{\sf}
\newtheorem{lem}{Lemma}[section]
\newtheorem{prop}{Proposition}[section]
\newtheorem{thm}{Theorem}[section]

\newtheorem{cort}[thm]{Corollary}
\newtheorem{corl}[lem]{Corollary}
\theoremstyle{plain}
\theoremseparator{~--}
\theorembodyfont{\rm}
\theoremsymbol{\sf ---}
\newtheorem{exm}{Example}[section]
\theoremstyle{nonumberplain}
\theoremseparator{~:}
\theoremsymbol{$\square$}
\newtheorem{prf}{Proof}

\begin{document}

\title{Enumerating planar locally finite Cayley graphs}
\author{David {\sc Renault}\footnote{Laboratoire Bordelais de Recherche en
Informatique, 351, cours de la Lib\'eration, Universit\'e Bordeaux I,
Talence, France -- Email: \texttt{renault@labri.fr}}}  

\maketitle

\begin{abstract}
We characterize the set of planar locally finite Cayley graphs, and
give a finite representation of these graphs by a special kind of
finite state automata called labeling schemes. As a result, we are
able to enumerate and describe all planar locally finite Cayley graphs
of a given degree. This analysis allows us to solve the problem of
decision of the locally finite planarity for a word-problem-decidable
presentation.

\medskip \noindent \textbf{Keywords:} {vertex-transitive, Cayley
graph, planar graph, tiling, labeling scheme}
\end{abstract}




\addcontentsline{toc}{section}{Introduction}
\section*{Introduction}

Given a group $G$ and a set of generators $A$, the associated Cayley
graph is a natural representation of the action of the generators on
$G$. When a Cayley graph is planar, its embedding possesses symmetry
properties induced by the choice of the geometry. The planarity
hypothesis also casts a new light on many decidability problems on
groups that are unsolvable in the general case. 

The problem of enumerating planar Cayley graphs goes back to Maschke
\cite{Maschke}, who enumerated in 1896 all finite groups having planar
Cayley graphs, these groups corresponding to the finite subgroups of
the symmetry group of the sphere. Nevertheless, for a given group $G$,
having a planar Cayley graph is a Markov property \cite{LyndonSchupp},
hence there is no algorithm deciding whether a finitely generated
group possesses a presentation for which its Cayley graph is planar or
not.

Presentations of groups having planar Cayley graphs have already been
discussed by Levinson \cite{LevinsonWPS}, who describes the geometry
of his graphs in terms of point and weak-point symmetry and gives
necessary conditions on the presentations of these groups. More recent
results by Droms et al. \cite{DromsCP,DromsPCG} develop the links
between connectivity and planarity inside Cayley graphs, which give
hints about the structure of the group. Considering the broader
problem of describing the vertex-transitive planar graphs, Imrich and
Fleischner \cite{Fleischner} solved the question in the case of finite
graphs. 

In this paper, we give an exhaustive description of the class of the
locally finite planar Cayley graphs, and answer some decidability
problems. Our results are based on Chaboud's work \cite{Chaboud}, who
studied the planar Cayley graphs that were also normal tilings. We
represent Cayley graphs by finite state automata called {\it labeling
schemes} along with a local geometrical invariant called a {\it type
vector}. We show that there exists a bijection between this
representation and the class of locally finite planar Cayley
graphs. This extends Chaboud's work by providing a more general class
of graphs. We give algorithmic means to describe each of these Cayley
graphs, along with their presentations. Our main result
(p.~\pageref{thm:enum}) is~:

\medskip


\setcounter{section}{4}
\setcounter{thm}{3}
\begin{thm}[Enumeration] 
Given $n \geq 2$, it is possible to effectively enumerate all
planar locally finite Cayley graphs having internal degree $n$, along
with one representative presentation. 
\end{thm}
\setcounter{section}{0}
\setcounter{thm}{0}

\medskip

Each Cayley graph belonging to this class is effectively computable
({\it i.e.} there exists an algorithm able to build any finite ball of
the graph), which leads to algorithms solving the word problem. When
the word problem is already decidable in a given presentation, it is
possible to decide whether the corresponding Cayley graph is
planar locally finite or not. These graphs also correspond to tilings
of the plane (seen as spherical, Euclidean or hyperbolic) by regular
polygons, and their automorphisms are realized by isometries.
Our techniques are similar to those used for groups acting on planar
surfaces \cite{Wilkie,Zieschang}.

\tableofcontents

\newpage

\section{Definitions}

A {\it directed graph} $\Gamma$ consists of a pair $(V,E)$, $V$ being
the set of {\it vertices} and $E \subset V \times V$ the set of {\it
edges}. Each edge corresponds to a pair of vertices $(s_1,s_2)$, $s_1$
being the {\it initial} vertex and $s_2$ the {\it terminal} one. A
{\it labeling} of the graph is an application from the set of edges
into a finite set $L$ of labels. The {\it degree} of a vertex is the
number of edges incident to this vertex. A {\it morphism} from the
graph $\Gamma_1 = (V_1,E_1)$ into $\Gamma_2 = (V_2,E_2)$ is an
application $\sigma :V_1\rightarrow V_2$ that preserves the edges of
the graphs. When both graphs are labeled, we impose that the morphisms
also preserve the labels of the edges. A {\it path} of $\Gamma$ is a
sequence of edges $(e_n)$ in $\Gamma$ such that $\forall n$, the
terminal vertex of $e_n$ and the initial vertex of $e_{n+1}$ are the
same. A {\it cycle} is a finite path whose initial and terminal
vertices are the same. A {\it loop} is a cycle containing a single
edge.

Let $G$ be a finitely generated group, and $A = \{a_1,\dots a_d\}$ a
set of generators of $G$. The {\it Cayley graph} $\Gamma$ of $G$ with
respect to this set of generators is the directed graph defined as
follows : the vertices of $\Gamma$ are the elements of $G$, and we
draw an edge labeled by a generator $a_i$ from a vertex $g_1$ to a
vertex $g_2$ whenever $g_1 \cdot a_i = g_2$ in the group $G$. 
We shall only be concerned with the case where the graph contains no
loops. Therefore, we prevent $A$ from containing the identity of $G$.
We impose that $A$ be stable by inversion in $G$. Thus, to any edge
linking $s_1$ to $s_2$ and labeled by $a_i$, there exists a reciprocal
edge from $s_2$ to $s_1$ labeled by $a_i^{-1}$. To avoid redundancy,
we reduce both edges into a single undirected edge linking $s_1$ to
$s_2$. 

A graph is said to be {\it planar} if it can be embedded in the plane,
without edges crossing or intersecting other vertices than their
extremities. Such an embedding is said to be {\it locally finite} if
its vertices have no accumulation point in the plane -- equivalently
there exists at most one accumulation point in the sphere. Our
embeddings will be considered {\it tame}, meaning that all edges are
$\mathcal{C}^1$ images of $[0;1]$. A graph is said to be {\it
vertex-transitive} if and only if, given any two vertices $(s_1,s_2)
\in \Gamma$, there exists an automorphism of $\Gamma$ mapping $s_1$
onto $s_2$. If $\Gamma$ is the Cayley graph of a group
$G$, then it is vertex-transitive.

A graph is {\it connected} if, for every pair of vertices $(s_1,s_2)$
of the graph, there exists a finite path in the graph with extremities
$s_1$ and $s_2$. A {\it n-separation} is a set of $n$ vertices whose
removal separates the graph in two or more connected components. A
{\it cut-vertex} of $\Gamma$ is a $1$-separation of $\Gamma$. A graph
is $n$-{\it separable} if it contains a $n$-separation. If it contains
no $n$-separation, it is {\it (n+1)-connected}. A graph is {\it
uniform} when all its vertices have the same degree $d$. The graphs we
will be dealing with are connected and uniform. There exists only one
non-trivial Cayley graph of degree $1$, which is the Cayley graph of
$\Rel/2\Rel$, corresponding to a single edge. Cayley graphs of degree
$2$ corresponds to cyclic groups $\Rel/n\Rel$ where $n$ might be
infinite. Thus, we shall only be interested in graphs of degree $d
\geq 3$.

Given a specific embedding of a locally finite planar graph, a {\it
face} $\mathcal{F}$ is defined as an arc-connected component of the
complement of the graph in the plane. $\mathcal{F}$ is said to be
{\it finite} when it is bounded in the plane, otherwise it is said to
be {\it infinite}. The {\it border} of this face, noted
$\partial\mathcal{F}$, is its boundary in topological terms.

\section{Properties of the locally finite planar graphs}

In the following, $\Gamma$ will be a locally finite, planar,
connected, vertex-transitive graph. Thus we will speak of vertices,
edges and even faces of $\Gamma$, as defined above.
Our major concern when dealing with such a graph is the following :
having set aside the possibility of accumulation points of the
vertices, which properties remain when we take into account the whole
embedding ? The following approach explores the geometric properties
of our graphs.  

\begin{lem}[Edge locally finite]
Let $K$ be a compact set of the plane. It is possible to modify the
embedding of $\Gamma$ such that the number of edges of $\Gamma$
intersecting $K$ is finite. 
\end{lem}

\begin{prf}
Let $K$ be a compact set of the plane, intersecting infinitely many
edges of the embedding of $\Gamma$. Since this set is compact, it
meets only a finite number of vertices of $\Gamma$, and since the
graph is of finite degree, only a finite number of edges intersecting
$K$ are incident to vertices effectively belonging to $K$.

\begin{center}
\begin{pspicture}(-2,-1.2)(6,1.2)

\pscustom[linestyle=none,fillstyle=solid,fillcolor=notwhite]{
\pscircle[linestyle=dashed,linecolor=gray](0,0){1.1cm}
}
\pscircle[linestyle=dashed,linecolor=gray](0,0){1.1cm}

\psframe[fillstyle=solid,fillcolor=notwhite,linestyle=dashed,linecolor=gray]%
(3.5,-1)(4.5,-0.5)
\rput[l](4.7,-0.75){Compact $K$}

\pnode(3.5,0){Z1}
\cnode(4,0){2pt}{Z2}	\ncline{Z1}{Z2}
\pnode(4.5,0){Z3}	\ncline{Z2}{Z3}
\rput[l](4.7,0){Vertices in $K$}

\pnode(3.5,0.75){Y1}
\pnode(4.5,0.75){Y3}	\ncline[linewidth=0.3pt]{Y1}{Y3}
\rput[l](4.7,0.75){Edges crossing $K$}

\cnode(0,0.4){2pt}{A}
\cnode(-1.1,0.7){2pt}{A1}	\ncline{A}{A1}
\cnode(1.1,0.7){2pt}{A2}	\ncline{A}{A2}	
\cnode(0.2,0.9){2pt}{A3}	\ncline{A}{A3}
\cnode(-0.2,0.9){2pt}{A4}	\ncline{A}{A4}
\cnode(-0.7,1.1){2pt}{A5}	\ncline{A1}{A5}\ncline{A5}{A4}

\cnode(0,-0.4){2pt}{B}
\cnode(-1.1,-0.7){2pt}{B1}	\ncline{B}{B1}
\cnode(1.1,-0.7){2pt}{B2}	\ncline{B}{B2}	
\cnode(0.2,-0.9){2pt}{B3}	\ncline{B}{B3}
\cnode(-0.2,-0.9){2pt}{B4}	\ncline{B}{B4}
\cnode(0,-1.2){2pt}{B5}		\ncline{B3}{B5}\ncline{B4}{B5}

\psset{linewidth=0.3pt}
\pscurve(-2.7,0.6)(0,0.3)(2.7,0.6)
\pscurve(-2.8,0.4)(0,0.2)(2.8,0.4)
\pscurve(-2.9,0.2)(0,0.1)(2.9,0.2)
\pscurve(-3,0.1)(0,0.05)(3,0.1)
\pscurve(-2.7,-0.6)(0,-0.3)(2.7,-0.6)
\pscurve(-2.8,-0.4)(0,-0.2)(2.8,-0.4)
\pscurve(-2.9,-0.2)(0,-0.1)(2.9,-0.2)
\pscurve(-3,-0.1)(0,-0.05)(3,-0.1)
\pscurve[linestyle=dashed](-3.1,0)(0,0)(3.1,0)

\end{pspicture}
\end{center} 

Consider now the connected components of the graph restricted to $K$
(we consider only the edges and vertices entirely belonging to
$K$). If there is just one component, then it is possible to modify
the embedding by pushing the accumulating edges outside of the
compact, while preserving the locally finite aspect of the graph. If
there is more than one component, then consider the subset of edges
intersecting $K$ but not incident to vertices of $K$, each edge
corresponding to a couple of vertices $(s,t)$. Since the graph is
locally finite, it is possible to extract from this set a series of
edges $(s_n,t_n)_{n\in\Nat}$ such that $d(s_n,K) \rightarrow \infty$
and $d(t_n,K) \rightarrow \infty$.  Thus the graph can not be
connected, since there is no way of joining the two components without
crossing an infinite number of edges belonging to the series
$(s_n,t_n)$, which is impossible.
\end{prf}

Following this lemma, it is possible to construct an {\it edge-locally
finite} embedding of $\Gamma$ in the following sense : by considering
$K_0 \subset \dots \subset K_n \subset \dots$ a series of compacts
sets of the plane such that every point of the plane eventually falls
into a $K_n$, and pushing all accumulation points towards infinity. This
property allows us to consider that the embedding of $\Gamma$ 
contains no accumulation points of the vertices. For such an
embedding, the set of points of $\Gamma$, adding the point at infinity
if $\Gamma$ is an infinite graph, is a closed subset of the sphere.
Faces of $\Gamma$ are therefore connected open subsets of the sphere.
From now on, we will consider that the embedding possesses this
property.

\begin{lem}[Non-intersection of faces] \label{lem:separation}
The border of a face of $\Gamma$ has no self-intersection. 
\end{lem}

\begin{prf}
Assume on the contrary that $\Gamma$ possesses a vertex $s$ belonging
at least twice to the border of a given face $\mathcal{F}$. There
exists a ball $B$ centered on $s$ such that $B\cap\mathcal{F}$
possessed more than one connected component. And there exists a path drawn
in the face connecting one of these components to the other. In
concrete terms, the situation looks like the following :

\begin{center}
\begin{pspicture}(-2,-1)(6,1)
\psset{unit=0.9}

\pscustom[linestyle=none,fillstyle=solid,fillcolor=notwhite]{%
	\pspolygon(-1.2,0.7)(0,0)(1.2,0.7)(1.8,0.4)(2.4,0.8)(2.1,-0.2)%
	(2.2,-0.8)(1.2,-0.7)(0,0)(-1.2,-0.7)(-1.2,-1.2)(-1,-1.4)(3,-1.4)%
	(3.2,-1.2)(3.2,1.2)(3,1.4)(-1,1.4)(-1.2,1.2)}
\pscustom[linestyle=none,fillstyle=solid,fillcolor=lightgray]{%
	\pspolygon(0,0)(1.2,0.7)(1.8,0.4)(2.4,0.8)(2.1,-0.2)%
	(2.2,-0.8)(1.2,-0.7)(0,0)}
\pscurve[linestyle=dashed,linecolor=gray](0,0)(0.5,0.8)(1.5,1.3)(3,0.4)(3,-0.4)%
	(1.5,-1.3)(0.5,-0.8)(0,0)

\cnode(0,0){2pt}{A}
\rput(0,-0.4){$s$}

\cnode(1.2,0.7){2pt}{C1}	\ncline{A}{C1}
\cnode(1.2,-0.7){2pt}{C2}	\ncline{A}{C2}
\cnode(1.8,0.4){2pt}{C3}	\ncline{C1}{C3}
\cnode(1.3,0){2pt}{C4}		\ncline{C1}{C4} 
\cnode(2.1,-0.2){2pt}{C5}	\ncline{C2}{C4}	\ncline{C4}{C5}
\pnode(2.4,0.8){D1}		\ncline[linestyle=dashed]{C3}{D1}
\pnode(2.2,-0.8){D2}		\ncline[linestyle=dashed]{C2}{D2}
\ncline{C3}{C5}			\ncline[linestyle=dashed]{C5}{D1}
				\ncline[linestyle=dashed]{C5}{D2}

\cnode(-1.2,-0.7){2pt}{C1}	\ncline{A}{C1}
\cnode(-1.2,0.7){2pt}{C2}	\ncline{A}{C2}
\cnode(-1.8,-0.4){2pt}{C3}	\ncline{C1}{C3}
\cnode(-1.3,0){2pt}{C4}		\ncline{C1}{C4}
\cnode(-2.1,0.2){2pt}{C5}	\ncline{C2}{C4}	\ncline{C4}{C5}
\pnode(-2.4,-0.8){D1}		\ncline[linestyle=dashed]{C3}{D1}
\pnode(-2.2,0.8){D2}		\ncline[linestyle=dashed]{C2}{D2}

\rput(0.75,0){$\Psi$}
\rput(-0.25,1){$\mathcal{F}$}

\psframe[fillstyle=solid,fillcolor=notwhite,linestyle=none]%
(3.75,-0.15)(4.75,-0.65)
\rput[l](4.95,-0.4){Face $\mathcal{F}$ and path}
\pnode(3.75,-0.4){Y1}
\pnode(4.75,-0.4){Y3}	\ncline[linestyle=dashed,linecolor=gray]{Y1}{Y3}

\psframe[fillstyle=solid,fillcolor=lightgray,linestyle=solid]%
(3.75,0.15)(4.75,0.65)
\rput[l](4.95,0.4){Subgraph $\Psi$}

\end{pspicture}
\end{center}

Since $\Gamma$ is locally finite, we consider the finite subgraph
$\Psi$ of $\Gamma$ contained in one finite connected component of the
plane separated by the path drawn inside the face.  Call $\varphi$ an
automorphism of $\Gamma$ mapping $s$ on a vertex of $\Psi$ different
from $s$. Since $s$ must be a cut-vertex of $\Gamma$, $\varphi(s)$ is
a cut-vertex of $\varphi(\Gamma)$, which separates $\Psi$ into two
distinct components, one of which must be finite and entirely inside
$\Psi$, otherwise $\Psi$ would be infinite. This component then
corresponds to $\varphi(\Psi)$. But this means that the number of
vertices in $\Psi$ is strictly greater than the number of vertices of
$\varphi(\Psi)$, which is impossible.
\end{prf}

The preceding lemma implies that around a given vertex of $\Gamma$,
and a face incident to that vertex, there exists at most two edges
incident both to the vertex and to the face. Let's consider more
closely the borders of the faces. 
Our embeddings are considered to be {\it tame}, meaning that all edges
are $\mathcal{C}^1$ images of the unit segment. Tame embeddings possess
the following sructure property :

\begin{lem}[Border structure] \label{lem:border}
Let $\mathcal{F}$ be a face of $\Gamma$. The border of $\mathcal{F}$
is either a finite cycle of $\Gamma$ or a two-way infinite path of
$\Gamma$ and its limit point at infinity. In particular, if
$(s_1,s_2)$ are any two vertices belonging to this border, then there
exists a unique finite path in the border with extremities $s_1$ and
$s_2$.
\end{lem}

\begin{prf}
Consider a face $\mathcal{F}$ of $\Gamma$ and a point $x \in
\partial\mathcal{F}$ that is not at infinity. Then $x$ is either on a
vertex or on an edge of $\Gamma$. We build a non self-intersecting
path going through $x$ if $x$ belongs to an edge, or going through the
two edges incident to $\mathcal{F}$ if $x$ is a vertex. Thanks to the
previous lemma, there exists either a cycle or a two-way infinite path
in $\Gamma$ possessing this property. In the second case, the locally
finite property implies that the extremities of the path must join at
infinity. In either alternative, this path constitues a simple closed
continuous curve of the sphere. According to the Jordan's theorem, it
divides the sphere in exactly two connected components, the curve
being the complete frontier of both parts. One of these components,
for example $\mathcal{H}$, must contain $\mathcal{F}$.

Consider a ball $B$ centered on $x$ such that $B$ only meets the edges
incident to $x$. Therefore, from lemma~\ref{lem:separation},
$\mathcal{H}\cap B = \mathcal{F}\cap B$. This implies that around $x$,
the border of $\mathcal{F}$ is the border of $\mathcal{H}$, that is to
say the path in $\Gamma$ going through $x$. Applying this property to
every point belonging to $\partial\mathcal{F}$, the border of
$\mathcal{F}$ must entirely contain the edges it is incident to, and
any vertex on its border is incident to two edges in the border. This
proves the lemma.
\end{prf}

This confirms the following natural intuition : the border of a finite
face is composed of a finite number of edges, whereas it is infinite
in the case of an infinite face.

\begin{lem}[Rule of intersection of the faces] \label{lem:intersect}
The intersection of two different faces of $\Gamma$, when not
empty, is either a single vertex, or an edge along with its
extremities. 
\end{lem}

\begin{prf}
Consider $\mathcal{F}_1$ and $\mathcal{F}_2$ two distinct faces of
$\Gamma$. Suppose that $\partial\mathcal{F}_1 \cap
\partial\mathcal{F}_2$ contains two distinct vertices $s_A$ and $s_B$
of $\Gamma$. Following lemma~\ref{lem:border}, these two vertices
$s_A$ and $s_B$ can be joined by a path in $\partial\mathcal{F}_1$,
and by a path in $\partial\mathcal{F}_2$. If these paths are the same,
then they must be reduced to a single edge because the degree of the
graph is $\geq 3$. Otherwise, we impose that the finite region
enclosed by the reunion of this paths does not contain either
$\mathcal{F}_1$ or $\mathcal{F}_2$. Concretely, the disposition of the
graph is the following~:

\begin{center}
\begin{pspicture}(-2,-1.25)(6.5,1.35)

\psset{linecolor=lightgray,hatchcolor=lightgray}

\pscustom[linestyle=none,fillstyle=solid,fillcolor=notwhite]{%
\pscurve[linestyle=dashed](-1.5,0.5)(-1,0.9)%
(-0.5,1.1)(0,0.8)(0.5,1)(1,0.9)(1.5,0.5)
\psline(1.5,0.5)(2,0)
\psline(2,0)(1.5,-0.5)
\pscurve[linestyle=dashed](1.5,-0.5)(1,-0.9)%
(0.5,-1.1)(0,-0.8)(-0.5,-1)(-1,-0.9)(-1.5,-0.5)
\psline(-1.5,-0.5)(-2,0)
\psline(-2,0)(-1.5,0.5)}
\pscustom[linestyle=none,fillstyle=vlines]{%
\pscurve[linestyle=dashed](-1.5,0.5)(-1,0.9)%
(-0.5,1.1)(0,0.8)(0.5,1)(1,0.9)(1.5,0.5)
\psline(1.5,0.5)(2,0)
\psline(2,0)(3,1.3)
\psline(3,1.3)(-3,1.3)
\psline(-3,1.3)(-2,0)
\psline(-2,0)(-1.5,0.5)}
\pscustom[linestyle=none,fillstyle=hlines]{%
\pscurve[linestyle=dashed](-1.5,-0.5)(-1,-0.9)%
(-0.5,-1)(0,-0.8)(0.5,-1.1)(1,-0.9)(1.5,-0.5)
\psline(1.5,-0.5)(2,0)
\psline(2,0)(3,-1.3)
\psline(3,-1.3)(-3,-1.3)
\psline(-3,-1.3)(-2,0)
\psline(-2,0)(-1.5,-0.5)}

\psline(2,0)(3,1.3)
\psline(2,0)(3,-1.3)
\psline(-2,0)(-3,1.3)
\psline(-2,0)(-3,-1.3)
\psset{linecolor=black}
\cnode(-1.5,0.5){2pt}{A1}	
\cnode(1.5,0.5){2pt}{A3}
\cnode(-2,0){2pt}{S1}		\ncline[linestyle=solid]{S1}{A1}
\cnode(2,0){2pt}{S2}		\ncline[linestyle=solid]{A3}{S2}
\cnode(-1.5,-0.5){2pt}{B1}	\ncline[linestyle=solid]{S1}{B1}
\cnode(1.5,-0.5){2pt}{B3}	\ncline[linestyle=solid]{B3}{S2}
\rput*(-2.4,0){$s_A$}
\rput*(2.4,0){$s_B$}
\rput*(-2,0.9){$\mathcal{F}_1$}
\rput*(-2,-0.9){$\mathcal{F}_2$}
\pscurve[linestyle=dashed](-1.5,0.5)(-1,0.9)%
(-0.5,1.1)(0,0.8)(0.5,1)(1,0.9)(1.5,0.5)
\pscurve[linestyle=dashed](-1.5,-0.5)(-1,-0.9)%
(-0.5,-1)(0,-0.8)(0.5,-1.1)(1,-0.9)(1.5,-0.5)
\psline[linestyle=dotted](-2,0)(-2.7,-0.5)
\psline[linestyle=dotted](-2,0)(-2.7,0.5)
\psline[linestyle=dotted](-2,0)(-1.2,-0.3)
\psline[linestyle=dotted](-2,0)(-1.2,0.3)
\psline[linestyle=dotted](-2,0)(-0.8,-0.05)
\cnode(-0.8,-0.05){1pt}{DD}
\rput(-0.8,0.1){$s$}
\psline[linestyle=dotted](2,0)(2.7,-0.5)
\psline[linestyle=dotted](2,0)(2.7,0.5)
\psline[linestyle=dotted](2,0)(1.2,-0.3)
\psline[linestyle=dotted](2,0)(1.2,0.3)
\psline[linestyle=dotted](2,0)(0.8,-0.05)
\psframe[fillstyle=hlines,linecolor=lightgray](4,0.5)(5,1.)
\rput[l](5.2,0.75){Face $\mathcal{F}_1$}
\psframe[fillstyle=solid,fillcolor=notwhite](4,-0.25)(5,0.25)
\rput[l](5.2,0){Internal part $\Phi$}
\psframe[fillstyle=vlines,linecolor=lightgray](4,-0.5)(5,-1)
\rput[l](5.2,-0.75){Face $\mathcal{F}_2$}
\end{pspicture}
\end{center}

Hence the cycle is a true Jordan curve in the plane. The gray part of
the picture corresponds to the finite connected component of the plane
separated by this curve. Call $\Phi$ the finite subgraph of $\Gamma$
inside this component.  We proceed by induction on the number of
vertices in $\Phi$. If $\Phi$ contains only two distinct vertices,
then the result is true. Consider now a vertex $s\in\Phi$, different
from $s_A$ and $s_B$, and call $\varphi$ the automorphism of $\Gamma$
mapping $s_A$ upon $s$.

Let's examine $\varphi(\Phi)$. $\Phi \subsetneq \varphi(\Phi)$ is
impossible, because $\Phi$ and $\varphi(\Phi)$ have the same number of
vertices. Likewise, $\varphi(\Phi) \subsetneq \Phi$.  Suppose {\it ab
absurdo} that $\Phi=\varphi(\Phi)$. There exists a neighbor of $s_A$
in $\Gamma$ not in $\Phi$, otherwise $\mathcal{F}_1 =
\mathcal{F}_2$. This neighbor, is mapped by $\varphi$ onto a vertex at
distance $1$ from $s$, which means a vertex of $\Phi$. Since
$\varphi_{|\Phi}$ is a bijection, this is absurd.  Therefore,
$\varphi(\Phi)$ intersects $\Phi$ and $\Gamma \backslash \Phi$.

Take $t \in \Gamma$ a vertex both in $\varphi(\Phi)$ and $\Gamma
\backslash \Phi$. Since $\Phi$ is connected, consider a simple path in
$\varphi(\Phi)$ joining $s$ and $t$. Since $s_A$ and $s_B$ are the
only cut-vertices between $\Phi$ and $\Gamma \backslash \Phi$, this
path must contain at least one of them. It can't contain both of them,
otherwise, when removing $\Phi$ from $\Gamma$, you obtain one or two
connected components, while removing $\varphi(\Phi)$ from $\Gamma$,
you obtain two or three connected components. Therefore, without loss
of generality, if $s_B$ is the only cut-vertex in $\varphi(\Phi)$,
then every path from $s$ to $t$ contains $s_B$. Every path from
$s_A$ to $\varphi^{-1}(t)$ contains $\varphi^{-1}(s_B)$. In
particular, $\varphi^{-1}(s_B)$ is a cut-vertex for $\Phi$, which is
absurd under our hypothesis. By induction, these parts are edges of
$\Gamma$. Since the degree of $\Gamma$ is $\geq 3$, it is impossible
to obtain such a thing.
\end{prf}

\begin{corl} \label{facetoface}
The automorphisms of $\Gamma$ map finite faces of $\Gamma$ into finite
faces of $\Gamma$.
\end{corl}

\begin{prf}
Consider the border of a finite face $\mathcal{F}$. Let $\varphi$ be an
automorphism of $\Gamma$ mapping $\partial\mathcal{F}$ on a cycle of
$\Gamma$ of the same length. This cycle defines a Jordan curve of the
plane. Let $\Phi$ be the finite subgraph in the finite connected
component delimited by this cycle, we examine
$\varphi^{-1}(\Phi$). This subgraph is connected to
$\partial\mathcal{F}$, so it intersects $\mathcal{F}$. Since it can't
be connected to another vertex in $\Gamma$ than the vertices incident
to $\mathcal{F}$, it must lie inside another face of $\Gamma$, namely
$\mathcal{H}$. Following the preceding lemma, the intersection of
$\mathcal{F}$ and $\mathcal{H}$ is entirely in the border of
$\mathcal{F}$. Hence $\Phi=\emptyset$~.
\end{prf}

\section{Rotation and inversion}

The geometrical properties of planar graphs are often affected by
their connectivity. For example, 3-connected planar graphs have a
strong geometric property : they possess a unique embedding in the
plane\cite{Whitney} or equivalently their dual is uniquely defined.
Chaboud~\cite{Chaboud} studied graphs possessing normality properties,
which in fact correspond to 3-connectivity and local finiteness. The
connectivity of the graph is directly linked to the number of infinite
faces around a vertex :

\begin{lem}[Infinite faces]
$\Gamma$ is $2$-separable $\Leftrightarrow$ each vertex is incident to
a unique infinite face. 
$\Gamma$ is $1$-separable $\Leftrightarrow$ each vertex is
incident to two or more infinite faces. 
\end{lem}

 In this section, we first recall the lemmas of Chaboud for 3-connected
graphs, and then extend his results in the complementary case of the
2-separable graphs.

\subsection{3-connected case}

In the 3-connected case, all faces of the graph are finite. Consider
any vertex $s$ of $\Gamma$ : as a definition, the generators appearing
around $s$ are the labels of the edges of terminal vertex $s$. The
planarity of $\Gamma$ imposes a cyclic order on these generators.

\begin{lem}[Chaboud's rotation lemma]
If $\Gamma$ is a Cayley graph with respect to a set of generators $A$,
then the generators appear in the same cyclic order around each
vertex, up to the direction of rotation.
\end{lem}

\begin{prf}
According to corollary~\ref{facetoface}, the faces incident to a given
vertex $s$ are mapped, by an automorphism $\varphi$, onto the faces
incident to the image vertex. Consider the set of faces incident to
$s$ as a finite set of cycles containing $s$, intersecting only on the
edges incident to $s$ :

\begin{center}
\begin{pspicture}(-2,-1.2)(2,1.2)

\cnode(-1.5,0){2pt}{A}
\cnode(-1.7,0.5){2pt}{A1}	\ncline{A}{A1}
\cnode(-2,0){2pt}{A2}		\ncline{A}{A2}
\cnode(-1.6,-0.6){2pt}{A3}	\ncline{A}{A3}
\cnode(-1.1,-0.3){2pt}{A4}	\ncline{A}{A4}
\cnode(-1,0.4){2pt}{A5}		\ncline{A}{A5}

\pscurve[linestyle=dashed,linecolor=gray](-1.7,0.5)(-2.5,0.8)(-2,0)
\pscurve[linestyle=dashed,linecolor=gray](-2,0)(-2.5,-0.8)(-1.6,-0.6)
\pscurve[linestyle=dashed,linecolor=gray](-1.6,-0.6)(-1,-1.1)(-1.1,-0.3)
\pscurve[linestyle=dashed,linecolor=gray](-1.1,-0.3)(-0.6,0)(-1,0.4)
\pscurve[linestyle=dashed,linecolor=gray](-1,0.4)(-1.2,1)(-1.7,0.5)

\rput(0,0){\Large $\rightsquigarrow$}

\cnode(1.5,0){2pt}{B}
\cnode(1.7,0.5){2pt}{B1}	\ncline{B}{B1}
\cnode(2,0){2pt}{B2}		\ncline{B}{B2}
\cnode(1.6,-0.6){2pt}{B3}	\ncline{B}{B3}
\cnode(1.1,-0.3){2pt}{B4}	\ncline{B}{B4}
\cnode(1,0.4){2pt}{B5}		\ncline{B}{B5}

\pscurve[linestyle=dashed,linecolor=gray](1.7,0.5)(2.5,0.1)(1.6,-0.6)
\pscurve[linestyle=dashed,linecolor=gray](2,0)(1.9,-1)(1.1,-0.3)

\pnode(2.05,-0.4){D}
\pnode(3,-1){E}	\nccurve[linecolor=red,angleA=180,angleB=320]{->}{E}{D}
\rput(3.2,-1){$w$}

\end{pspicture}
\end{center}

Since faces are mapped onto faces by automorphisms of the graph, then
adjacent faces are mapped onto adjacent faces. Therefore, the borders
of the images of the faces incident to $s$ can't cross, which implies
that the order of rotation of the generators is the same. 
\end{prf}

\begin{lem}[Chaboud's inversion lemma]
If an edge labeled by a generator $a_i$ joins two vertices of
different orientations, then it is the case for every edge labeled by
$a_i$.
\end{lem}

\begin{prf}
The proof is essentially the same as in the preceding lemma, but
instead of considering the paths of the border of the faces incident
to a vertex, we consider an edge $e$ labeled by $a_1$, and the two
paths corresponding to the borders of the faces containing $e$.

\begin{center}
\begin{pspicture}(-2,-1.2)(2,1.2)
\psset{unit=0.8}

\cnode(-2.5,-0.75){2pt}{A}
\cnode(-2.5,0.75){2pt}{B}		\ncline{A}{B}
\cnode(-3.25,1){2pt}{B1}		\ncline{B1}{B}
\cnode(-1.75,1){2pt}{B2}		\ncline{B2}{B}
\cnode(-3.25,-1){2pt}{A1}		\ncline{A1}{A}
\cnode(-1.75,-1){2pt}{A2}		\ncline{A2}{A}

\nccurve[linecolor=gray,linestyle=dashed,angleA=150,angleB=210]{A1}{B1}
\nccurve[linecolor=gray,linestyle=dashed,angleA=30,angleB=330]{A2}{B2}

\rput(-2.2,0){$a_1$}
\rput(-2.1,1.1){$a_2$}
\rput(-2.9,1.1){$a_d$}
\rput(-2.1,-1.1){$a_i$}
\rput(-2.9,-1.1){$a_j$}

\rput(0,0){\Large $\rightsquigarrow$}

\cnode(2.5,-0.75){2pt}{C}
\cnode(2.5,0.75){2pt}{D}		\ncline{C}{D}
\cnode(3.25,1){2pt}{D1}			\ncline{D1}{D}
\cnode(1.75,1){2pt}{D2}			\ncline{D2}{D}
\cnode(3.25,-1){2pt}{C1}		\ncline{C1}{C}
\cnode(1.75,-1){2pt}{C2}		\ncline{C2}{C}

\rput(2.8,0){$a_1$}
\rput(2.1,1.1){$a_d$}
\rput(2.9,1.1){$a_2$}
\rput(2.1,-1.1){$a_i$}
\rput(2.9,-1.1){$a_j$}

\pscurve[linecolor=gray,linestyle=dashed]%
(1.75,1)(1.25,-1.25)(2.75,-1.25)(3.25,-1)
\pscurve[linecolor=gray,linestyle=dashed]%
(3.25,1)(3.75,-1.25)(2.25,-1.25)(1.75,-1)

\pnode(2.5,-1.3){F}
\pnode(3.5,-1.75){G}	\nccurve[linecolor=red,angleA=180,angleB=320]{->}{G}{F}
\rput(3.7,-1.75){$w$}

\end{pspicture}
\end{center}

Since these paths do not cross before the automorphism, they can't
cross after having been mapped. Therefore if the generator labeling
$e$ inverses the direction of rotation of the generator, it is also
the case of all edges labeled by $a_1$. 
\end{prf}

A generator $a_i$ is said to be {\it direct} if it always joins
vertices of the same orientation. It is said to be {\it indirect} in
the other case. As a simple corollary, we have that : 

\begin{lem}[Type Vector]
Let $s$ be a vertex of $\Gamma$, and
$\{\mathcal{F}_1,\dots,\mathcal{F}_d\}$ the set of faces of $\Gamma$
incident to $s$, ordered in cyclic order around $s$. Let
$\mathcal{V}=[k_1,\dots,k_d]$ where $k_i$ is the number of edges
incident to $\mathcal{F}_i$. Then $\mathcal{V}$ is independent of the
vertex $s$, up to a cyclic permutation or reversal of the vector.
\end{lem}

This vector is called the {\it type vector} of the graph. Not all
possible type vectors correspond to a locally finite graph
$\Gamma$ : \cite{Tilings} describes the special case of the Euclidean
plane, where among the 21 possible type vectors for tilings by regular
polygons, only 12 correspond to vertex-transitive graphs, whereas the
remaining 9 can't possibly describe a vertex-transitive
graph. Moreover, to a given type vector may correspond multiple
non-isomorphic vertex-transitive graphs, or even Cayley graphs, as is
the case for $[4,4,4,6]$ :

\begin{center}
\includegraphics[width=3.2cm]{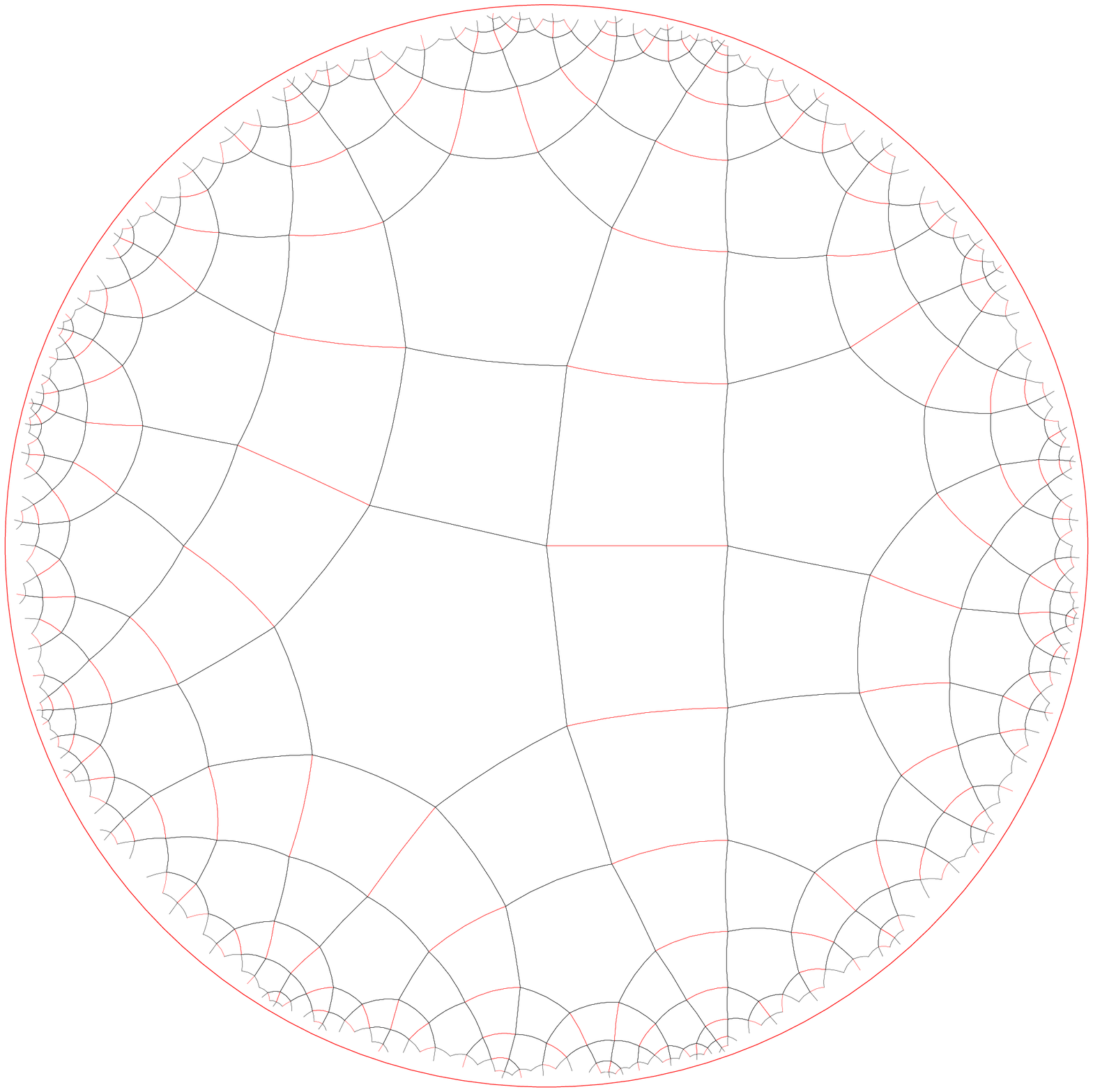}
\hskip 1.5cm
\includegraphics[width=3.2cm]{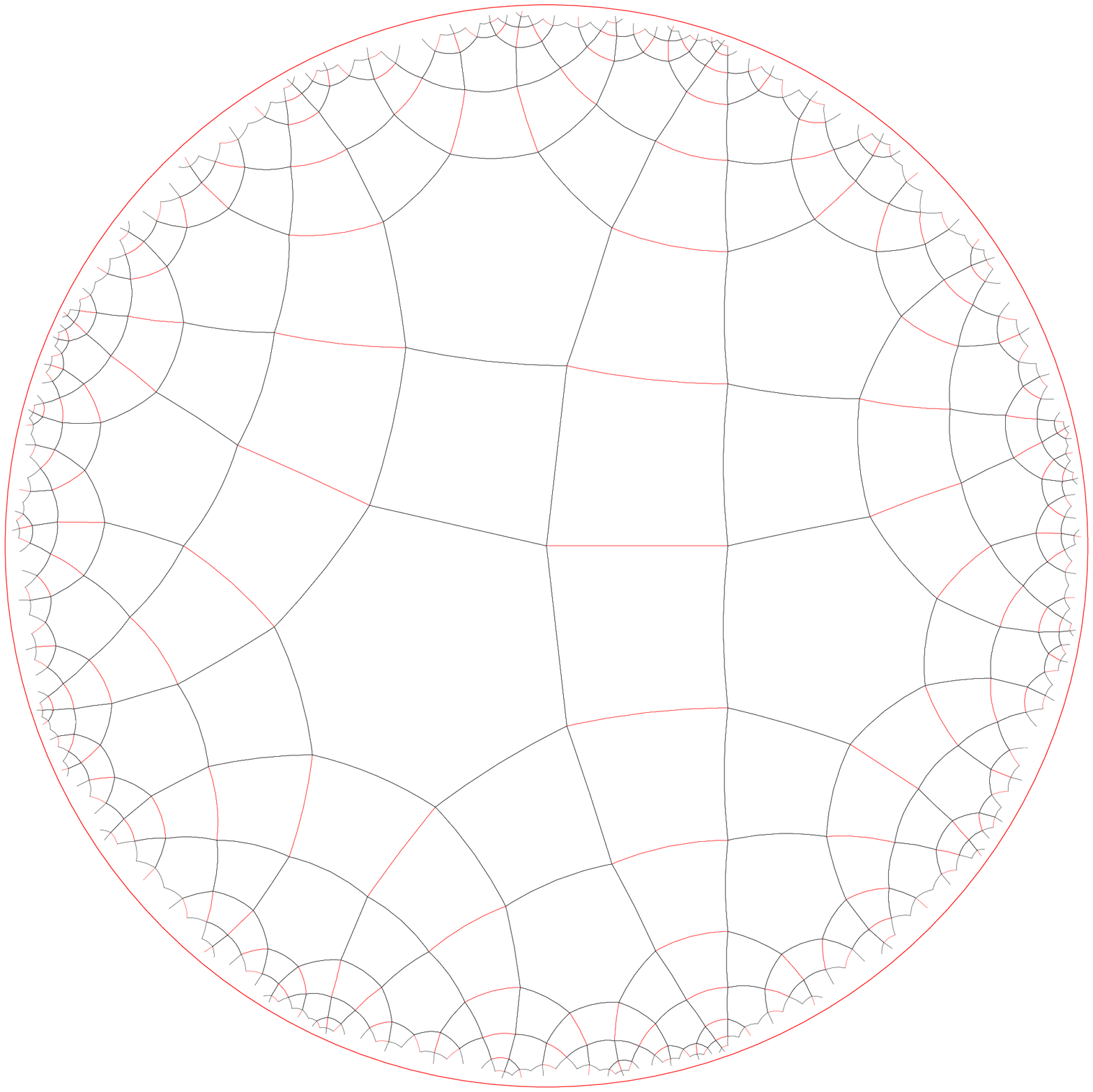}
\end{center}

\subsection{2-separable case}

In the case of connectivity lower than 3, infinite faces appear. For
instance, if we consider the Cayley graphs of free groups (infinite
regular trees), one can easily devise embeddings that violate the
preceding lemmas.  A graph $\Gamma$ is said to possess a {\it proper
labeling} or to be {\it weakly point-symmetric} \cite{LevinsonWPS} if,
for a specific embedding, it verifies the rotation lemma and the
inversion lemma. The previous subsection shows that 3-connected graphs
must have a proper labeling, but this is not general. Our purpose in this
section is to prove that, nevertheless, there always exists an
embedding of the graph that follows the lemmas of rotation and
inversion.

\medskip

Let $\Gamma$ be a $2$-separable, locally finite, planar and connected
vertex-transitive graph. If any vertex of $\Gamma$ is a cut-vertex,
then every vertex is a cut-vertex. If we slit open the graph along its
cut-vertices, the remaining connected components are called {\it
2-connected components} or CC$_2$.  Since the graph is
vertex-transitive, the set of components incident to a vertex is
independent of the vertex, and finite. These components are not
necessarily vertex-transitive themselves, and they may be finite or
infinite. By hypothesis, 2-connected components are locally finite
graphs. 

\noindent \parbox{8.3cm}{\parfillskip=0pt Suppose now that $\Gamma$
is a Cayley graph. Two generators $a_i$ and $a_j$ are said to be
equivalent ($a_i\sim a_j$) when the pair of edges incident to a vertex
and labeled by $a_i$ and $a_j$ belong to the same CC$_2$. This is an
equivalence relation on the set of generators $A$. If a generator is
alone in its class, then the CC$_2$ consists only of a single
edge. Let's consider an equivalence class $\mathcal{C}$ for this}
~
\lower 1.3cm \hbox{{\begin{pspicture}(-1.65,-1.5)(1.85,1.5)
\psset{unit=0.7}

\pscustom[fillstyle=solid,fillcolor=notwhite,linestyle=none]{%
\pspolygon(0,0)(1,1.7)(1.9,1.7)(2.3,1.0)(2,0)(2.3,-1.0)(1.9,-1.7)(1,-1.7)}
\pscustom[fillstyle=solid,fillcolor=notwhite,linestyle=none]{%
\pspolygon(0,0)(-1.2,1.6)(-2,0)}

\cnode(0,0){2pt}{A}
\cnode(1,1.7){2pt}{B1}	\ncline{A}{B1} 	\lput{:U}{\rput[c]{U}(0,.25){$a_1$}}
\cnode(2,0){2pt}{B2}	\ncline{A}{B2} 	\lput{:U}{\rput[c]{U}(0,.25){$a_2$}}
\cnode(1,-1.7){2pt}{B3}	\ncline{A}{B3} 	\lput{:U}{\rput[c]{U}(0,.25){$a_3$}}

\cnode(-1.2,1.6){2pt}{D1}\ncline{A}{D1}	\lput{:D}{\rput[c]{U}(0,.25){$a_4$}}
\cnode(-2,0){2pt}{D2}	\ncline{A}{D2}	\lput{:D}{\rput[c]{U}(0,.25){$a_5$}}
\ncline{D1}{D2}
\cnode(-1.2,-1.6){2pt}{E}\ncline{A}{E}	\lput{:D}{\rput[c]{U}(0,.25){$a_6$}}

\cnode(1.9,1.7){2pt}{C1}
\cnode(2.3,1.0){2pt}{C2}	\ncline{B1}{C1}	\ncline{C1}{C2}\ncline{C2}{B2}
\cnode(1.9,-1.7){2pt}{C3}
\cnode(2.3,-1.0){2pt}{C4}	\ncline{B3}{C3}
\ncline{C3}{C4}\ncline{C4}{B2}

\psline[linecolor=lightred,linestyle=dashed](0,-2)(0,-0.5)
\psline[linecolor=lightred,linestyle=dashed](0,2)(0,0.5)
\psline[linecolor=lightred,linestyle=dashed](-0.35,-0.15)(-2.1,-0.9)
\end{pspicture}}}

\vskip 2pt
\noindent relation. If $s$ is a vertex of $\Gamma$, it is possible to join
the two vertices $s\cdot a_i$ and $s\cdot a_j$ by a path remaining in
the component corresponding to $\mathcal{C}$ and avoiding the vertex
$s$. The number of classes is equal to the number of CC$_2$ appearing
around a vertex. If $\Gamma$ is $2$-connected, then it possesses a
unique CC$_2$ equals to~$\Gamma$.

\begin{center}
{\begin{pspicture}(-2,-1.6)(8,1.6)
\psset{unit=0.7}

\pscustom[fillstyle=solid,fillcolor=notwhite,linestyle=none]{%
\pspolygon(0,0)(1,1.7)(1.9,1.7)(2.3,1.0)(2,0)(2.3,-1.0)(1.9,-1.7)(1,-1.7)}
\pscustom[fillstyle=solid,fillcolor=notwhite,linestyle=none]{%
\pspolygon(0,0)(-1.2,1.6)(-2,0)}

\cnode(0,0){2pt}{A}
\cnode(1,1.7){2pt}{B1}	\ncline{A}{B1} 	\lput{:U}{\rput[c]{U}(0,.25){$a_1$}}
\cnode(2,0){2pt}{B2}	\ncline{A}{B2} 	\lput{:U}{\rput[c]{U}(0,.25){$a_2$}}
\cnode(1,-1.7){2pt}{B3}	\ncline{A}{B3} 	\lput{:U}{\rput[c]{U}(0,.25){$a_3$}}

\cnode(-1.2,1.6){2pt}{D1}\ncline{A}{D1}	\lput{:D}{\rput[c]{U}(0,.25){$a_4$}}
\cnode(-2,0){2pt}{D2}	\ncline{A}{D2}	\lput{:D}{\rput[c]{U}(0,.25){$a_5$}}
\ncline{D1}{D2}
\cnode(-1.2,-1.6){2pt}{E}\ncline{A}{E}	\lput{:D}{\rput[c]{U}(0,.25){$a_6$}}

\cnode(1.9,1.7){2pt}{C1}
\cnode(2.3,1.0){2pt}{C2}	\ncline{B1}{C1}	\ncline{C1}{C2}\ncline{C2}{B2}
\cnode(1.9,-1.7){2pt}{C3}
\cnode(2.3,-1.0){2pt}{C4}	\ncline{B3}{C3}
\ncline{C3}{C4}\ncline{C4}{B2}

\psline[linecolor=lightred,linestyle=dashed](0,-2)(0,-0.5)
\psline[linecolor=lightred,linestyle=dashed](0,2)(0,0.5)
\psline[linecolor=lightred,linestyle=dashed](-0.35,-0.15)(-2.1,-0.9)

\rput(4,-0.3){\LARGE $\rightsquigarrow$}
\rput(4,0.3){\small automorphism}

\pscustom[fillstyle=solid,fillcolor=notwhite,linestyle=none]{%
\pspolygon(8,0)(9,1.7)(9.9,1.7)(10.3,1.0)(10,0)(10.3,-1.0)(9.9,-1.7)(9,-1.7)}
\pscustom[fillstyle=solid,fillcolor=notwhite,linestyle=none]{%
\pspolygon(8,0)(6.8,-1.6)(6,0)}

\cnode(8,0){2pt}{A}
\cnode(9,1.7){2pt}{B1}	\ncline{A}{B1} 	\lput{:U}{\rput[c]{U}(0,.25){$a_3$}}
\cnode(10,0){2pt}{B2}	\ncline{A}{B2} 	\lput{:U}{\rput[c]{U}(0,.25){$a_2$}}
\cnode(9,-1.7){2pt}{B3}	\ncline{A}{B3} 	\lput{:U}{\rput[c]{U}(0,.25){$a_1$}}

\cnode(6.8,-1.6){2pt}{D1}\ncline{A}{D1}	\lput{:D}{\rput[c]{U}(0,.25){$a_5$}}
\cnode(6,0){2pt}{D2}	\ncline{A}{D2}	\lput{:D}{\rput[c]{U}(0,.25){$a_4$}}
\ncline{D1}{D2}
\cnode(6.8,1.6){2pt}{E}\ncline{A}{E}	\lput{:D}{\rput[c]{U}(0,.25){$a_6$}}

\cnode(9.9,1.7){2pt}{C1}
\cnode(10.3,1.0){2pt}{C2}	\ncline{B1}{C1}	\ncline{C1}{C2}\ncline{C2}{B2}
\cnode(9.9,-1.7){2pt}{C3}
\cnode(10.3,-1.0){2pt}{C4}	\ncline{B3}{C3}
\ncline{C3}{C4}\ncline{C4}{B2}

\psline[linecolor=lightred,linestyle=dashed](8,-2)(8,-0.5)
\psline[linecolor=lightred,linestyle=dashed](8,2)(8,0.5)
\psline[linecolor=lightred,linestyle=dashed](7.65,0.2)(5.9,1.2)
\end{pspicture}}
\end{center} 

As in the 3-connected case, the order of the generators around a
vertex in a given class is independent of the vertex, allowing for the
two directions of rotation. The automorphisms of $\Gamma$ may only
modify (a) the cyclic order of the CC$_2$ appearing around a given
vertex and (b) apply a symmetry to a given CC$_2$. Let's consider the
set of CC$_2$ around a vertex, namely
$\{\mathcal{C}_1,\dots,\mathcal{C}_p\}$. This vertex defines a cyclic
order on the set of generators. Our purpose consists in building an
embedding of the graph such that the generators appear in the same
cyclic order around each vertex.

\begin{exm} 

\parfillskip=0pt The following graph is an example of 1-separable
Cayley graph of degree 5. Its sole 2-connected component, up to
isomorphism, is composed

\noindent \hbox{\parbox{8.3cm}{ \parfillskip=0pt plus 1 fil of two
triangles joined 
by an edge. Notice that two of these components appear around each
vertex, and that they are not equivalent : one component has three
edges incident to the vertex, while the other has only two. The
automorphisms of the graph may modify the ordering of the generators
around a given vertex by twisting these components. A possible
presentation for this group is :}} 
~ \lower 1.37cm
\hbox{\includegraphics[width=3.5cm]{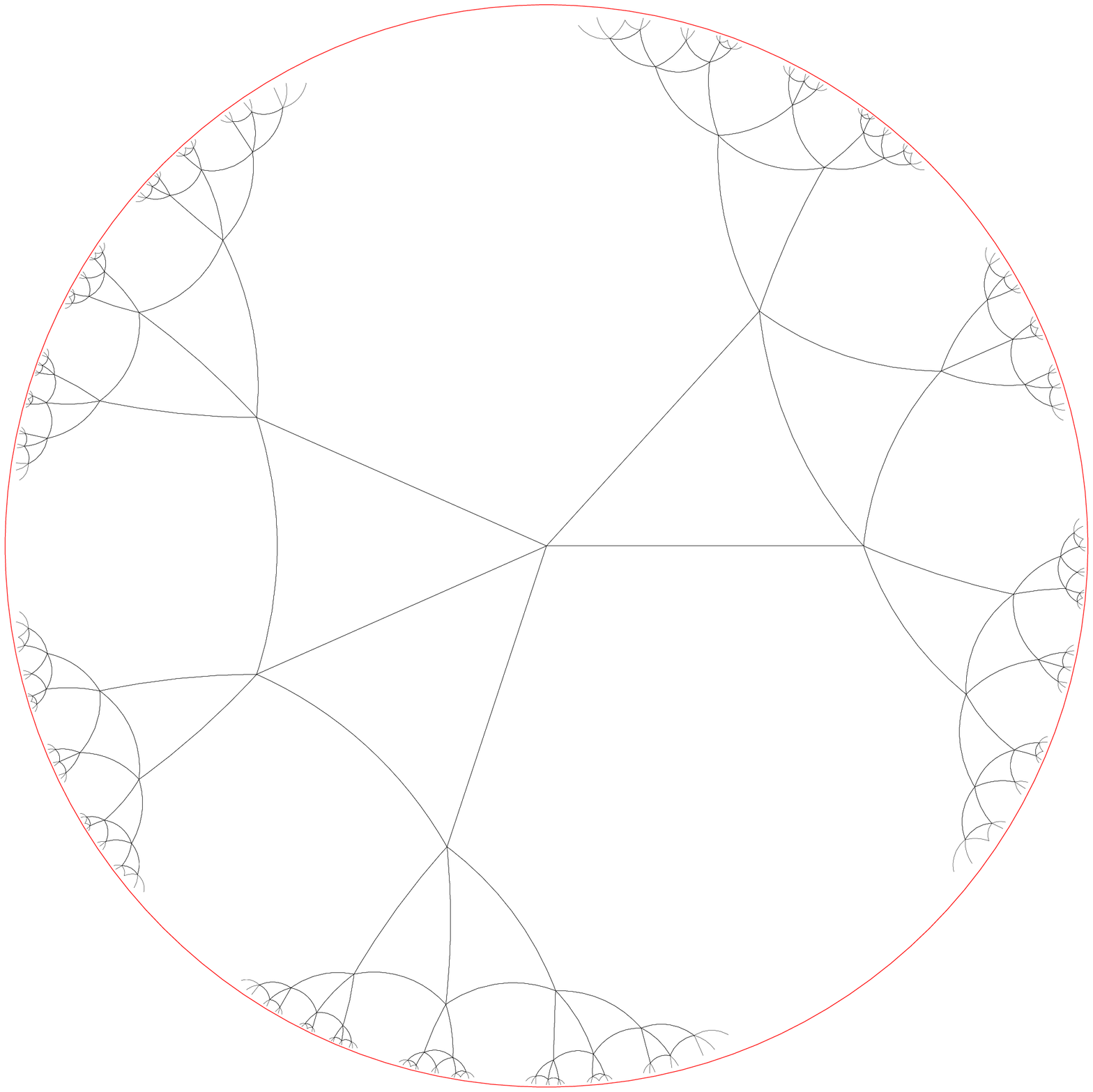}}
$$ G = \bigg\langle a_1,\dots,a_5 ~\bigg|~ 
a_4^2,a_2a_3,a_1a_5,a_3a_4a_1 \bigg\rangle  $$
\end{exm} 
\parfillskip=0pt plus 1.0fil

\begin{prop}[Proper labeling] \label{labeling}
There exists an embedding of $\Gamma$ that obeys the inversion and
rotation lemmas. Therefore the graph $\Gamma$ possesses a proper
labeling of its edges. 
\end{prop}

\begin{prf}
Consider a vertex $s$ of $\Gamma$, and define with this vertex an
ordering of the generators of the group. Let's build an embedding of
$\Gamma$ by induction : at the $n$-th step of the induction, the graph
$\Gamma_n$ is a locally finite graph, isomorphic to a subgraph of
$\Gamma$, such that every vertex in $\Gamma$ at distance less than $n$
from $s$ belongs to this subgraph. We also impose that the distance in
the plane between a vertex and $s$ be superior to the distance in
$\Gamma$. 

Let's choose for $s$ a point in the plane. Let $ \mathcal{C} =
\{\mathcal{C}_1,\dots,\mathcal{C}_p\}$ be the set of CC$_2$ around a
vertex in $\Gamma$. By symmetry and rearrangement of the CC$_2$, it is
possible to define an order on $\mathcal{C}$ which is coherent with
the order of the generators in each CC$_2$. Each CC$_2$ being locally
finite and independent in $\Gamma$ of the others, it is possible to
embed each one of them such that the distance condition be verified,
and that the order of the generators around $s$ be correct, up to a
cyclic permutation. We call this graph $\Gamma_1$.

Suppose now that we have built $\Gamma_n$, and consider the finite set
of vertices at distance $n+1$ around which we have not built all the
necessary CC$_2$ yet. Around each vertex in this set, we have only
built one CC$_2$. If this CC$_2$ is reduced to a single edge, then we
choose the order of the generators around the vertex to be the same
that the order around the other extremity. Otherwise, the order is
already defined by the order of the generators in the CC$_2$.

Let's consider the face incident to our vertex that does not belong to
the component already attached to the vertex. From lemma~\ref{lem:border},
the border of this face is either finite, or a bi-infinite path whose
only accumulation point is at the infinity. In each case, it is always
possible to embed the missing CC$_2$ in this face, such that the graph
still verifies the distance condition and that the order of the
generators around $s$ is correct. After having added a finite number
of components, all the vertices at distance $n+1$ are completed, and
we obtain~$\Gamma_{n+1}$. 

The limit graph is isomorphic to $\Gamma$. Since the distance
condition is verified, it is locally finite, and by construction, it
obeys the rotation lemma. During the construction, we have chosen the
generators alone in their equivalence class to be direct. All the
others follow the inversion lemma by an argument similar to Chaboud's
lemmas.  Hence, this embedding obeys the inversion lemma, as long as
the rotation lemma.
\end{prf}

\noindent \hbox{\parbox{8.3cm}{\parfillskip=0pt For such a graph, it is
therefore possible to extend the notion of type vector that we defined
in the 3-connected case. Around a vertex, we may have finite or
infinite faces. And considering the embedding provided by the previous
proposition, the automorphisms of $\Gamma$ also map infinite faces
onto infinite faces. Therefore we allow infinite faces in type
vectors, the type vector of $\Gamma$ being independent of the chosen
vertex. For example,}}~ \lower 1.5cm 
\hbox{\includegraphics[width=3.5cm]{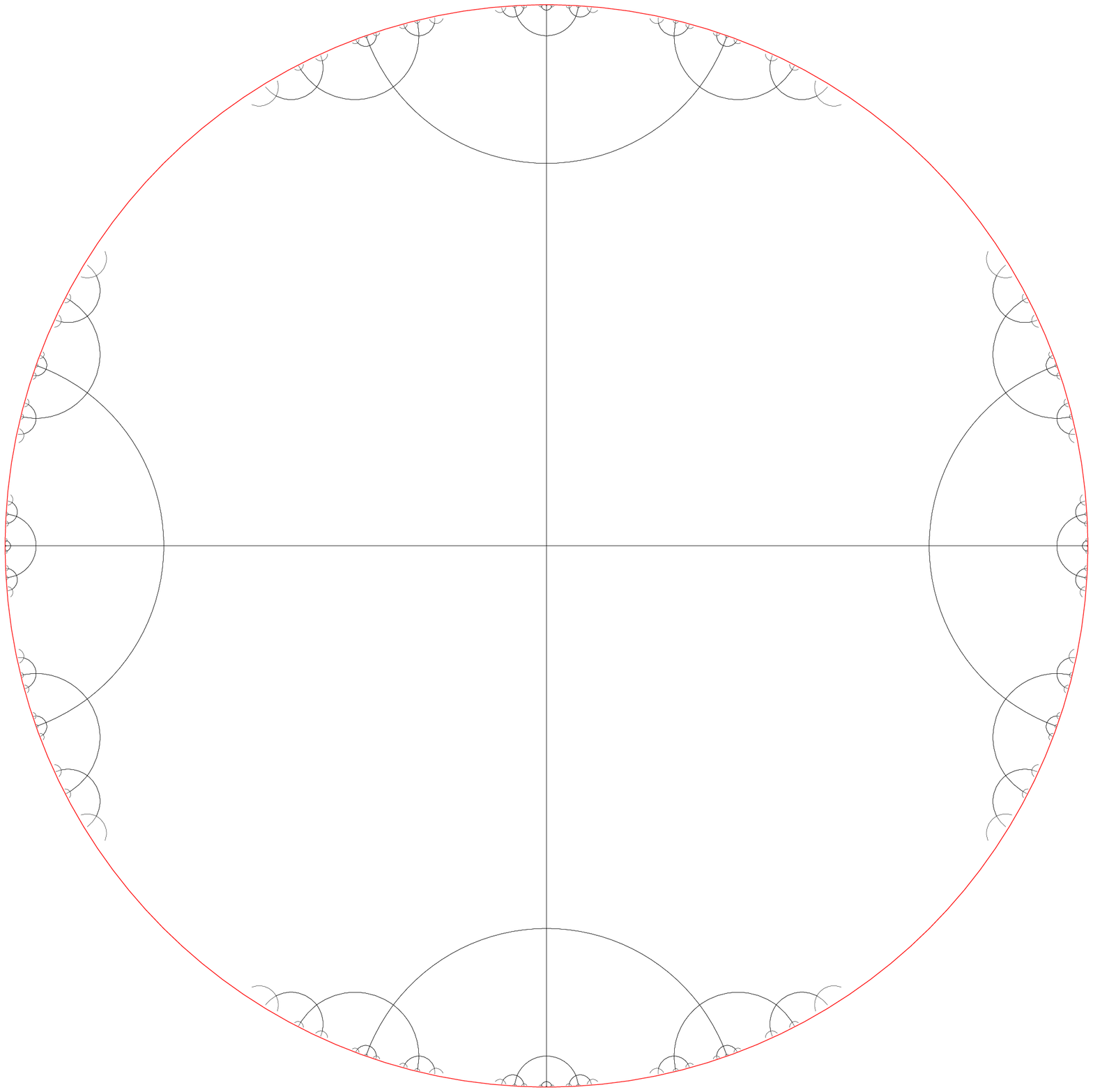}}

\vskip 2pt \noindent 
the free group on two elements, with the usual generators, possesses a
planar Cayley graph of degree $4$, and its type vector is
$[\infty,\infty,\infty,\infty]$.


\section{Labeling schemes}

\subsection{Definitions and properties}

Let $A=\{a_1,\dots,a_d\}$ a set of generators. A {\it labeling
scheme} on $A$ is given by a 3-tuple $(p,\sigma,\tau)$ :

\begin{itemize}
\item a cyclic permutation $p$ on $A$;
\item an involution $\sigma$ on $A$ ($\sigma = \sigma^{-1}$);
\item a function $\tau:A\rightarrow\{-1;1\}$ such \\ that for
each $x\in A$, $\tau(\sigma(x)) = \tau(x)$.
\end{itemize}

The permutation $p$ corresponds to the cyclic ordering of the
generators around any vertex of $\Gamma$. The involution $\sigma$
links every generator with its inverse in $G$. The function $\tau$
assigns the value $1$ to a generator if it is direct, and $-1$ if it
is indirect. The previous section assures that every locally finite
planar Cayley graph possesses a labeling scheme that is coherent with
its embedding, {\it i.e.} describes the order of rotation of its
generators around any vertex and the direct/indirect property of the
generators. Since we will only be concerned with Cayley graphs, we can
choose $p=[1,2,\dots,d]$ by renaming the generators of the graph. 

Using the labeling scheme, it is possible to describe the border of
the faces with a finite state automaton, using the general ideas from
maps and hypermaps \cite{MachiCori}. Let's consider a vertex $s$ from
$\Gamma$ around which the generators appear in cyclic order along the
clockwise direction. If we follow the edge incident from $s$ labeled
by the generator $a_i$, we can compute the direction of rotation of
the generators around the terminal vertex of the edge. Then it is
possible to rebuild the border of any face of the graph, be it finite
or infinite. 

\begin{center}
\begin{pspicture}(-2,-1.5)(1.8,2.2)
\psset{unit=1.3}

\cnode(-1.75,1.25){2pt}{Y}
\cnode(-1.25,-0.75){2pt}{Z}
\cnode(-1,0.5){2pt}{A}		\ncline{->}{Z}{A}\ncline{->}{Y}{A}
\cnode(0,1.25){2pt}{B}		
\cnode(1.3,0.9){2pt}{C}		

\nccurve[angleA=55,angleB=200]{<-}{A}{B}
\nccurve[linestyle=dotted,angleA=10,angleB=250]{->}{A}{B}
\nccurve[angleA=15,angleB=140]{<-}{B}{C}
\nccurve[linestyle=dotted,angleA=330,angleB=180]{->}{B}{C}

\newcommand{\pp}{\textcolor{red} p}

\rput(-1,0.8){$s$}
\rput*(-1.1,-0.15){$a_1$}
\rput*(-1.35,0.9){$a_3$}
\rput*(-0.55,1.05){$a_2$}
\rput*(-0.3,0.65){$a_{\sigma(2)}$}
\rput*(0.75,1.3){$a_{\pp(\sigma(2))}$}
\rput*(1.1,0.6){$a_{\sigma(\pp(\sigma(2)))}$}

\rput(0,-1.2){Case were $a_2$ is direct}
\end{pspicture}
\hspace{2.5cm}
\begin{pspicture}(-2,-1.5)(1.8,2.4)
\psset{unit=1.3}

\cnode(-1.75,1.25){2pt}{Y}
\cnode(-1.25,-0.75){2pt}{Z}
\cnode(-1,0.5){2pt}{A}		\ncline{->}{Z}{A}\ncline{->}{Y}{A}
\cnode(0,1.25){2pt}{B}		
\cnode(1.3,0.9){2pt}{C}		

\nccurve[angleA=55,angleB=200]{<-}{A}{B}
\nccurve[linestyle=dotted,angleA=10,angleB=250]{->}{A}{B}
\nccurve[angleA=15,angleB=140]{<-}{B}{C}
\nccurve[linestyle=dotted,angleA=330,angleB=180]{->}{B}{C}

\newcommand{\pp}{\textcolor{red} p^{\textcolor{red} -\textcolor{red} 1}}

\rput(-1,0.8){$s$}
\rput*(-1.1,-0.15){$a_1$}
\rput*(-1.35,0.9){$a_3$}
\rput*(-0.55,1.05){$a_2$}
\rput*(-0.3,0.65){$a_{\sigma(2)}$}
\rput*(0.75,1.3){$a_{\pp(\sigma(2))}$}
\rput*(1.1,0.6){$a_{\sigma(\pp(\sigma(2)))}$}
\rput(0,-1.2){Case were $a_2$ is indirect}
\end{pspicture}

\end{center}

For a more formal point of view, let's consider the set of vertices $C
= C^+ \cup C^-$, where $C^+$ is composed of $d$ vertices
$\{1^+,\dots,d^+\}$ and $C^-$ is composed of $\{1^-,\dots,d^-\}$. Let
$s$ be a vertex of $C$. We define an action of $\Rel$ onto the set of
vertices of $C$ by~:
\begin{enumerate}
\item if $s^+ \in C^+$, then let $t = \sigma(p(s))$. \\
If $t$ is direct,
then $1 \cdot s^+ = t^+$, otherwise $1 \cdot s^+ = t^-$.
\item if $s^+ \in C^-$, then let $t = \sigma(p^{-1}(s))$. \\
If $t$ is direct,
then $1 \cdot s^- = t^-$, otherwise $1 \cdot s^- = t^+$.
\end{enumerate}
This action can be naturally extended into an action of
monoid. Moreover, it is invertible as a result of the properties of
the labeling scheme. Therefore it is an action of the group $\Rel$ on
a finite set. The orbits of this action correspond to the borders of
the faces, described by the generators labeling them, in function of
the direction of rotation of the vertices.  

It is possible to represent this action with a finite state automaton,
each state being a couple $(a_i,b)$ where $a_i \in A$ is a generator
and $b\in\{true,false\}$ being a boolean. Our alphabet is composed of
two different letters, $p$ for following to the next generator in the
current direction of rotation, and $i$ for reversing the generator :
\begin{itemize}
\item $(a_i,true) \overset{\longrightarrow}{p} (a_{(i+1 \mod d)},true)$ and 
 $(a_i,false) \overset{\longrightarrow}{p} (a_{(i-1 \mod
d)},false)$;
\item if $\tau(i)=1$, $(a_i,b) \overset{\longrightarrow}{i} (a_{\sigma(i)},b)$
otherwise $(a_i,b) \overset{\longrightarrow}{i} (a_{\sigma(i)},^\neg b)$.
\end{itemize}
Computing the border of a face consists in selecting a starting state
in the automaton, and reading the infinite word $(pi)^\omega$, thus
obtaining a finite cycle of the automaton. 

\begin{exm}
The following example has been taken from \cite{Chaboud}. The left
picture describes the labeling scheme as defined by Chaboud : the
cyclic permutation $p$ corresponds to a cyclic graph of order $8$, and
the involution $\sigma$ corresponds to the black arcs linking the
vertices of the cycle. A direct generator is pictured as a circle, an
indirect one as a square. 

\begin{center}
\begin{pspicture}(-2,-2.6)(2,2.6)
\SpecialCoor \degrees[8] 
\psset{nodesep=3pt}
\cnode(2;2){2pt}{A1}	
\pnode(2;3){A2}		\ncline{->}{A1}{A2}
\rput[c](2;3){\tiny $\square$} 
\cnode(2;4){2pt}{A3}	\ncline{->}{A2}{A3}
\cnode(2;5){2pt}{A4}	\ncline{->}{A3}{A4}
\cnode(2;6){2pt}{A5}	\ncline{->}{A4}{A5}
\pnode(2;7){A6}		\ncline{->}{A5}{A6}	
\rput[c](2;7){\tiny $\square$} 
\pnode(2;0){A7}		\ncline{->}{A6}{A7}	
\rput[c](2;0){\tiny $\square$}
\pnode(2;1){A8}		\ncline{->}{A7}{A8}	
\rput[c](2;1){\tiny $\square$}
			\ncline{->}{A8}{A1}

\nccurve[linewidth=1.5pt,angleA=1,angleB=3,offset=0pt]{A1}{A1}
\ncline[linewidth=1.5pt]{A2}{A7}
\nccurve[linewidth=1.5pt,angleA=3,angleB=5,offset=0pt]{A3}{A3}
\nccurve[linewidth=1.5pt,angleA=1,angleB=2]{A4}{A5}
\nccurve[linewidth=1.5pt,angleA=3,angleB=5]{A6}{A8}

\newcounter{iia}
\multido{\ia=2+1}{8}{%
\setcounter{iia}{\ia}
\addtocounter{iia}{-1}
\rput(2.4;\ia){\theiia}}
\end{pspicture}
~\hspace{2cm}~
\begin{pspicture}(-2,-2.7)(2,2.7)
\SpecialCoor \degrees[8] 

\multido{\ia=2+1}{8}{%
\setcounter{iia}{\ia}
\addtocounter{iia}{-1}
\rput(1.2;\ia){\footnotesize$\theiia^-$}
\pscircle(1.5;\ia){0.05}
\psline[arrows=->,arrowsize=5pt](1.5;\ia)(1.5;\theiia)
\rput(2.5;\ia){\footnotesize$\theiia^+$}
\psline[arrows=<-,arrowsize=5pt](2.1;\ia)(2.1;\theiia)
\pscircle(2.1;\ia){0.05}}
\psset{linecolor=lightgray,linewidth=1pt}
\pscurve(2.1;2)(2.3;2.1)(2.3;1.9)(2.1;2)
\pscurve(1.5;2)(1.7;2.1)(1.7;1.9)(1.5;2)
\pscurve(2.1;4)(2.3;4.1)(2.3;3.9)(2.1;4)
\pscurve(1.5;4)(1.7;4.1)(1.7;3.9)(1.5;4)
\pscurve(2.1;3)(1;2)(1.5;0)
\pscurve(1.5;3)(0.2;6)(2.1;0)
\pscurve(2.1;5)(2.3;5.5)(2.1;6)
\pscurve(1.5;5)(1.7;5.5)(1.5;6)
\pscurve(2.1;7)(1;0)(1.5;1)
\pscurve(1.5;7)(1;1.5)(2.1;1)

\psset{linecolor=red,linewidth=2pt}
\pscircle*(2.1;2){0.1}
\pscircle*(1.5;0){0.1}
\pscircle*(2.1;1){0.1}
\psset{linecolor=lightred}
\pscircle*(1.5;3){0.1}
\pscircle*(2.1;7){0.1}
\pscircle*(1.5;2){0.1}
\psset{linecolor=lightblue}
\pscircle*(2.1;3){0.1}
\pscircle*(2.1;4){0.1}
\pscircle*(2.1;6){0.1}
\pscircle*(1.5;1){0.1}
\psset{linecolor=lightgreen}
\pscircle*(1.5;4){0.1}
\pscircle*(1.5;5){0.1}
\pscircle*(1.5;7){0.1}
\pscircle*(2.1;0){0.1}
\psset{linecolor=gold}
\pscircle*(2.1;5){0.1}
\psset{linecolor=yellow}
\pscircle*(1.5;6){0.1}
\end{pspicture}
\end{center}

The picture on the right describes the action of $\Rel$ on the
generators. The action of $1 \in \Rel$ on a vertex is computed as
follows : follow the oriented black arc starting from the generator,
then follow the non-oriented gray arc to obtain the resulting
generator. By alternatively following black and gray arcs, one follows
the orbit of the generator.   
\end{exm}

Different orbits can correspond to same faces. This may happen if
$[a_1^{s_1} \dots a_k^{s_k}]$ and $[a_1^{-s_1} \dots a_k^{-s_k}]$ are
two different orbits, or if an orbit is composed of the inverses of
the generators of another orbit. Two such orbits are called {\it dual
orbits}. Dual orbits correspond to borders of the same face, but read
in the opposite direction. We say that two generators $a_i$ and $a_j$
{\it correspond to the same face} if $a_j^+$ belongs either to the
orbit of $a_i^+$ or to its dual orbit.

For a generator $a_i$ pointing toward the vertex $s$, let $k_i$ be
the smallest positive integer such that $k_i \cdot a_i^+ = a_i^+$. The
vector $[k_1,\dots,k_d]$ is called the {\it primitive vector} of the
labeling scheme. Another type vector $[l_1,\dots,l_d]$ is a {\it
valid type vector} with regard to this labeling scheme if and only
if : 

\begin{itemize}
\item $\forall i \in [ 1;d ], l_i$ is a
multiple of $k_i$;  
\item if $a_i^+$ and $a_j^+$ correspond to the same face, then
$l_i=l_j$.  
\end{itemize}

\begin{exm}
Continuing with the previous example, the orbits under the action of
$\Rel$ are : 

\begin{itemize}
\item $(~1^+~ 7^-~ 8^+~)$ with dual $(~6^+~2^-~ 1^-~)$, of
length $3$ ;
\item $(~2^+~ 3^+~ 5^+~ 8^-~)$ with dual $(~7^+~ 6^-~ 4^-~
3^-~)$, of length $4$;
\item $(~4^+~)$ with dual $(~5^-~)$ of length $1$.
\end{itemize}

Therefore the primitive type vector corresponding to the labeling
scheme is $[3,4,4,1,3,4,3]$. The case of length $1$ means that any
length is allowed for the face, assuming that it is greater than 3,
the smallest possible length for a face to exist. The generators 1, 8
and 6 belong to the same face, as is the case for 2, 3, 5 and 7. All
valid type vectors are then of the form $[3p,4q,4q,r,4q,3p,4q,3p]$,
where $(p,q,r)$ are positive integers, such that $r\geq 3$.
\end{exm}

We can extend this construction to type vectors with infinite
faces. We allow the number $\infty$ in the type vectors, to stand for
infinite faces. If $l_i = \infty$, then the first condition of
validity is always true. These type vectors can be seen as limits
of finite type vectors, as we shall see in the following sections.  

\subsection{Characterization}

Every locally finite planar Cayley graph possesses a labeling scheme,
and a type vector that is valid with regard to this scheme. We will
now take an interest in the converse of this result~: given a random
labeling scheme, and a valid type vector, is it possible to produce
a Cayley graph and the corresponding group that have the same labeling
scheme and type vector ? 

Our goal in this section consists in building specific embeddings of
Cayley graphs with particular geometrical properties. Depending on the
graph, we select an appropriate geometry : Euclidean, spherical or
hyperbolic. As long as our graphs are locally finite, embeddings in
spherical geometry will lead to finite graphs. Infinite faces thus can
only occur in Euclidean and hyperbolic geometry. The following theorem
constructs embeddings that are tilings of the plane by regular
polygons. A {\it regular polygon} is a non-intersecting cyclic graph
embedded in the plane such that all edges have the same length and the
angle between two consecutive edges is set. This allows for finite or
infinite polygons. In the Euclidean plane, an infinite regular polygon
corresponds to a half-plane, with border a straight line. In the
hyperbolic plane, there exists infinitely many non-isometric infinite
polygons\footnote{For example when the relation between the length $l$
of the edge and the angle $\theta$ between two edges is
$l=\frac{1}{2}\sqrt{2+2\cos\theta}$. In the Poincar\'e model, the
vertices of the border are the orbits of a parabolic isometry : they
belong to a circle tangent to the circle at infinity.}.

\begin{thm}[Existence] \label{primary}
Given a labeling scheme $(p,\sigma,\tau)$, and a valid type vector
$[k_1,\dots,k_d]$, there exists a locally finite planar Cayley graph
possessing this scheme and type vector. Moreover, all faces of the
embedding are regular polygons.
\end{thm}

\begin{prf}
The proof of this result involves two different parts : first, we show
it is possible to glue around a single vertex regular polygons
corresponding to the type vector. Then, we build the entire graph by
induction, as in proposition~\ref{labeling}. 

Consider any point in the plane. We are going to evaluate the interior
angle of a regular polygon of side length $l$, with $k_i$ sides, and
then the total angle corresponding to all the polygons in the type
vector must be equal to $2\pi$ :  
\renewcommand{\arraystretch}{1.5}
$$ 
\theta_i(l) = \left\{
\begin{array}{cl}
\textrm{Spherical :}&
\strut 2\arcsin \left( \frac{\cos(\pi/k_i)}{\cos(l/2)}\right) \\
\textrm{Euclidean :}&
\strut \frac{(k_i-2)}{k_i}\pi\\
\textrm{Hyperbolic :}&
\strut 2\arcsin \left( \frac{\cos(\pi/k_i)}{\cosh(l/2)}\right)
\end{array} 
\right. \quad \textrm{and} \quad \sum_{i=1}^d \theta_i(l) = 2\pi$$
This equation determines the choice of the geometry : let $\Sigma$ be
the sum of the angles in the Euclidean plane. If $\Sigma=2\pi$, we
choose the Euclidean geometry, and any value is possible for $l$ (the
corresponding graphs will be homothetic). If $\Sigma<2\pi$, we choose
the spherical geometry, whereas if $\Sigma>2\pi$ we select the
hyperbolic geometry. In any case, there exists an unique solution of
the equation\footnote{Except for the following type vectors :
$[3,3,p\geq 5]$, $[3,4,p\geq 6]$ and $[3,5,p\geq 9]$. There does not
exist a labeling scheme validating any of these type vectors (cf
p~\pageref{page:tv}).} that determines a unique value for the length $l$,
and consequently for the angles $\theta_i$. These values allow us to
draw all the regular polygons that correspond to our type vector
around a given point of the plane.

We will now build the graph by induction : let $\epsilon$ be a vertex
of $\Gamma$. The graph $\Gamma_n$ is a planar locally finite graph,
isomorphic to a subgraph of $\Gamma$, containing all vertices of
$\Gamma$ at distance $\leq n$ from $\epsilon$. The first graph
$\Gamma_1$ is composed of the glued polygons of the first
case. Suppose now that we have built $\Gamma_n$. Consider the finite
set of vertices of $\Gamma_n$ at distance $n$ for which all the
incident edges still have not been drawn. We shall build the remaining
edges with geodesics, such that the length of the geodesic be $l$,
and the angle at the vertex between the edge labeled by $a_i$ and the
edge labeled by $a_{i+1}$ be equal to $\theta_i(l)$.

The labeling scheme describes how the edges must be labeled : take a
vertex $s$ and an edge labeled by the generator $a_i$ pointing towards
$s$. Depending on the direction of rotation of the generators around
$s$, it is possible to build all edges incident to $s$, in such a way
that the angle between thed edge $a_i$ and the edge $a_{i+1}$ be
$\theta_i(l)$. The edge labeled by $a_i$ is seen by its other
extremity $t$ as an edge labeled by $a_{\sigma(i)}$ pointing towards
$t$. The direction of rotation of the generators around $t$ is
inverted if $\tau(i)=1$, otherwise it remains the same. On the same
principle, it is now possible to build the generators around $t$. 

\smallskip
\noindent \parbox{\textwidth}{\parfillskip=0pt Suppose {\it ab
absurdo} that this construction leads to two edges crossing. Then
consider one of the finite faces delimited by these two edges and the
remaining }

\noindent \hbox{\parbox{7.9cm}{\parfillskip=0pt part of
the graph. If there are no more edges to add into this face, then
since our construction fixes the angles between two edges, then this
angle is always the same inside the face. This comes from the fact
that the construction due to the labeling scheme follows the borders of the
faces. This face is a regular polygon, as the length and angles }}~
\lower 1.35cm \hbox{\begin{pspicture}(-3.7,-1.4)(0.2,1.5)

\pscustom[fillcolor=notwhite,fillstyle=solid,linestyle=none]{%
\pspolygon(0,0)(-1,1)(-2,1)(-3,0)(-2,-1)(-1,-1)}
\cnode(0,0){2pt}{A}	
\cnode(-1,1){2pt}{A1}	\ncline{->}{A}{A1}
\cnode(-2,1){2pt}{A2}	\ncline[linestyle=dotted]{->}{A1}{A2}
\cnode(-3.5,-0.5){2pt}{A3}	\ncline{->}{A2}{A3}
\cnode(-1,-1){2pt}{B1}	\ncline{->}{A}{B1}
\cnode(-2,-1){2pt}{B2}	\ncline[linestyle=dotted]{->}{B1}{B2}
\cnode(-3.5,0.5){2pt}{B3}	\ncline[linestyle=dashed]{->}{B2}{B3}
\rput(0.2,0){$e$}
\rput{320}(-0.3,0.7){$a_1$}
\rput{40}(-0.3,-0.7){$b_1$}
\rput{40}(-2.7,0.7){$a_n$}
\rput{320}(-2.7,-0.7){$b_m$}
\end{pspicture}}

\vskip 3.5pt \noindent have
been chosen in that sense. The only possible way the edges can then
cross is either at their extremities, or being in fact the same
edge. In each case, the face correspond to a regular polygon and the
edges do not really cross. In the case where there still exists edges
to add in the interior of the face, we reason by induction, adding the
finite set of remaining edges one by one.
\end{prf}

Moreover, this tiling of the plane is vertex-transitive : the
construction is independent of the starting vertex, and applying this
construction with two different starting vertices creates the
automorphism mapping the first vertex on the second. The labeling of
the edges entails that the automorphism group is simply transitive,
which imposes that the graph is a Cayley graph. It is possible to give
a canonical presentation of the underlying group : 

\begin{itemize}
\item Set of generators : $A=\{a_1,\dots,a_d\}$ (corresponding to the
free group of degree $d$);
\item Relators for the inverses : $a_ia_{\sigma(i)}$ for all
$i\in [ 1;d ]$;  

\item Relators for the border of the face between $a_i$ and $a_{i+1}$ : 
 since for any face, the length of the corresponding orbit
$\mathcal{O}_i = a_{j_1} a_{j_2} \dots a_{j_{p_i}}$ divides the length
of the face, the corresponding relator is $\mathcal{O}_i^{n_i}$ where
$n_i\cdot p_i = k_i$ is  the number in the type vector. For infinite
faces, there is no added relator. 
\end{itemize}

\begin{exm} \parfillskip=0pt The type vector $[3,4,4,3,4,3,4,3]$ is
validated by the previous labeling scheme. This leads us to a Cayley
graph of degree 8, pictured below.

\noindent \hbox{\parbox{8.3cm}{For this graph, $\Sigma = 14/3$,
consequently our embedding is built in the Poincar\'e model of the
hyperbolic plane. This graph has no infinite face, therefore it is
3-connected. The arrangement between the squares and the triangles is
directed by the labeling scheme. Following our method, it is possible
to compute any finite ball of the Cayley graph and embed it in the
hyperbolic plane with regular polygons. A presentation of the
underlying group is given by~: 
}}~ \lower 1.4cm \hbox{\includegraphics[width=3.5cm]{figure5.ps}}
$$ G = \bigg\langle a_1,\dots,a_8 ~\bigg|~ a_1a_6a_2,
a_7a_6a_4a_3,a_5^3, a_1^2,a_3^2,a_2a_7, a_6a_8, a_4a_5 
\bigg\rangle  $$ 
\end{exm}

\begin{exm} Finite planar Cayley graphs are embedded in the
sphere. 

\noindent \hbox{\parbox{8cm}{We find among them the Archimedean solids,
and graphs such as the one drawn on the right on the sphere, which is
called the {\it snub cube}. This one is of degree 5, and its type
vector is $[3,3,3,3,4]$. Square faces are glued such that no two of them
are adjacent. The corresponding group is of order 24, and a possible
presentation of this group is given by~: }}~
\lower 1.37cm \hbox{\includegraphics[width=3.5cm]{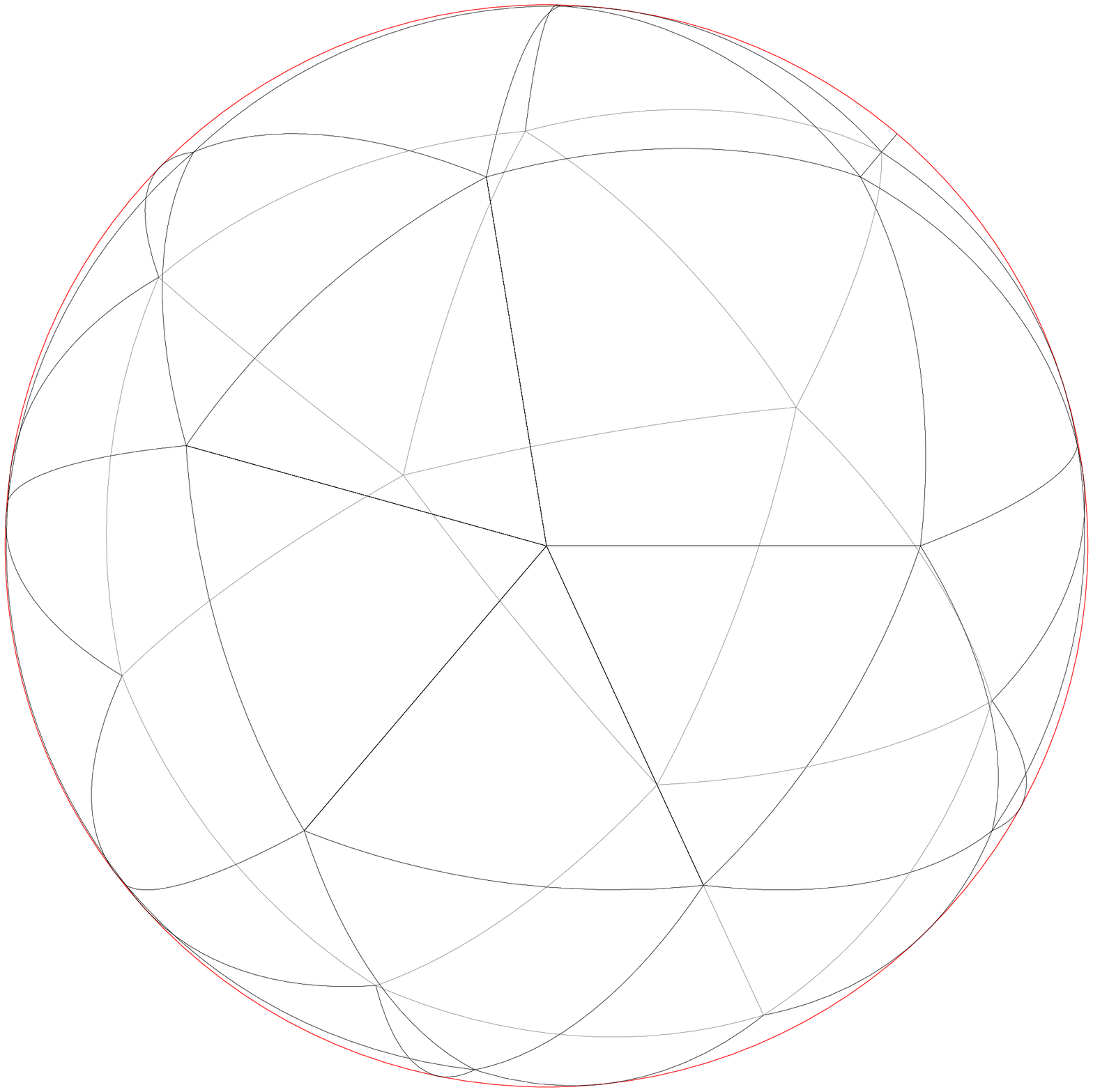}}

\vskip 1.5mm
$$ G = \bigg\langle a_1,\dots,a_5 ~\bigg|~ 
a_1a_2,a_3^2,a_4a_5,a_1^4,a_2a_3a_5,a_4^3 \bigg\rangle  $$  
\end{exm}

\begin{thm}[Coherence]
If $\Gamma$ is a locally finite Cayley graph that possesses a labeling
scheme $(p,\sigma,\tau)$ and a type vector $[k_1,\dots,k_d]$, then the
preceding construction yields a graph isomorphic to $\Gamma$.
\end{thm}

\begin{prf}
Consider $\Gamma$ our Cayley graph and $\Theta$ the graph produced by
the algorithm. These two graphs have the same sets of generators, and
share in common the relators corresponding to the border of the
faces. And every relator in $\Gamma$ correspond to a cycle in the graph,
that is to say a finite product of relators of the faces. Hence the
cycles of $\Gamma$ correspond to cycles of $\Theta$, and {\it vice
versa}. Therefore the two groups are the same, and because of the
uniqueness of the construction, the graphs are isomorphic.   
\end{prf}

Therefore, for any Cayley graph obeying our conditions, it is possible
to embed the graph in a particular geometry of the plane with regular
polygons. Moreover, automorphisms of the graph map faces onto
isometric faces, even infinite faces. Consequently, every automorphism
of $\Gamma$ extends into an isometry of the plane. Since the
automorphisms group of $\Gamma$ is isomorphic to the group $G$, then
$G$ is a discrete group of isometries of the plane, which sums up to~:

\begin{cort}
Every locally finite planar Cayley graph may be embedded in
the plane as a vertex-transitive tiling by regular
polygons. Automorphisms of the graph then correspond to isometries of
the plane. 
\end{cort} 

This result appears in \cite{BabaiGrowth} : since the tiling is made of
regular polygons, the growth of the graph is approximatively
equivalent to the growth of balls in the geometry. Therefore graphs
drawn in the Euclidean space are of quadratic growth, whereas graphs
drawn in the hyperbolic space are of exponential growth. 

\subsection{Combinatorial approach}

Consider at first the problem of the description of the set of
labeling schemes. Given a labeling scheme, it is simple to compute
its primitive vector. The number of different labeling schemes of
degree $d$ is bounded by $I(d)*2^d$, where $I(d)$ is the number of
involutions in $\mathfrak{S}_d$. It is therefore possible to express
the following corollary : 

\begin{thm}[Enumeration] \label{thm:enum}
Given a number $n \geq 2$, it is possible to effectively enumerate all
planar locally finite Cayley graphs having internal degree $n$, along
with their canonical presentation. 
\end{thm}

Having enumerated all possible labeling schemes of given degree, it
becomes possible to count all possible presentations that lead to
locally finite planar Cayley graphs. Another interest is, given a
planar vertex-transitive graph that is also locally finite, to test
whether this graph is a Cayley graph or not, and if the test is
positive, the list of group presentations corresponding to this
graph. It is already known that all archimedean tilings are Cayley
graphs. Yet the graph of the icosahedron in the sphere and the
$[5,5,5,5]$ isohedral tiling of the hyperbolic plane by pentagons are
both vertex-transitive graphs that are not Cayley. 

In terms of graph theory, it is also possible to enumerate all
primitive type vectors, for a given degree and, having removed the
redundant ones, to obtain all possible distinct vectors. For example,
the different primitive vectors of degree $3$ are given by the
following :  
$$ \label{page:tv} \bigg\{ [3n, 3n, 3n], [n, 2m, 2m], [2n, 2m, 2p]
\bigg\}$$

\dots for values of $n,m,p$ such that the faces of the graph are at
least triangles. The set of planar graphs obtained is strictly larger
than the set of hyperbolic graphs presented in \cite{Tilings}
({\it cf.} appendix). 
\medskip

Second, we devise algorithms to effectively describe the graph/tiling
corresponding to a given labeling scheme. The formulas given in
Theorem~\ref{primary} allow us to compute the angles and lengths of the
edges of the graph. Then it is possible to draw any finite subgraph in
the given geometry. In particular, given a word on the generators, it
is possible to compute approximately the position of the
corresponding element of the group in the tiling in linear time on the
length of the word. This leads to an algorithm to decide the word
problem in quadratic time. Our algorithms are used to draw in
Postscript language the graphs given in example in annex of this
article. 
\medskip

Third, we take interest in the following problem : given a finite
presentation of a group $G$, is it decidable whether this presentation
possesses a planar locally finite Cayley graph or not~? We impose that
the set of generators $A$ of the presentation be closed under
inversion. Such a presentation is said to be {\it full} if and only
if, if $l$ is the length of the longest relator in $R$, all elements
of $u\in A^\star$ of length $\leq l$ such that $u=\epsilon$ in $G$
belong to $R$. More precisely, the presentation contains all relators
of length $\leq l$. In the case of full presentations, the problem is
decidable : 

\begin{thm}[Full presentation $\Rightarrow$ Decidable]
Let $\langle A, R\rangle$ be a full presentation of the group
$G$. Then it is decidable whether this presentation possesses a
locally finite planar Cayley graph or not. 
\end{thm}

\begin{prf}
Suppose we possess a full presentation $\langle A, R \rangle$ of a
group $G$. Considering all relators of length $2$, we already know the
inverse vector $\sigma$, which maps $a_i$ on
$a_{\sigma(i)}=a_i^{-1}$. Consider $n$ the number of generators in
$A$. The number of labeling schemes possessing $n$ generators, and
having $\sigma$ as an inverse vector is finite, and the word problem
is decidable on any of the corresponding graphs. Call $\mathcal{A}$
this set schemes. It remains to find whether any of these
``candidates'' is the adequate Cayley graph for $G$. A necessary
condition consists in finding whether for a given candidate, the
borders of the faces belong to $G$ or not.

We state that in a full presentation, these borders must belong to
the presentation for the candidate to be valid. Suppose {\it ab
absurdo} that the border of one of the faces does not belong to the
presentation initially. Consider $l$ the size of the longest relator
in $R$. Either $l$ is greater than the size of the largest face in the
Cayley graph, and then it is decidable whether the relator belongs to
the presentation or not, or $l$ is smaller and it is impossible for
the relators in $R$ to generate the same group as the labeling scheme.

It suffices then to check if the remaining relators of the
presentation correspond to relators in $G$, which corresponds to the
resolution of the word problem. The exhaustion of all possible schemes
leads either to a negative result, or to a labeling scheme
corresponding to the initial presentation. 
\end{prf}

\begin{cort}[Word problem decidable case]
If the word problem is decidable on a presentation of $G$,
then it is decidable whether this presentation corresponds to a
locally finite planar Cayley graph or not.
\end{cort}

\addcontentsline{toc}{section}{Discussion}
\section*{Discussion}

The properties of locally finite planar Cayley graphs are rather
strong properties. There exists two distinct directions which could
enlarge our class or graphs. First, our study is restricted to Cayley
graphs, when the most part of graph theory deals with
vertex-transitive graphs. There exists graphs that are locally finite,
planar, vertex-transitive, but not Cayley graphs of groups. Studying
this class of graphs could lead to some intuition about the difference
between vertex-transitive and Cayley graphs, would it be only for
planar locally finite ones. Secondly, the most obvious property that
we could explore is the locally finite one. Our graphs contain only
one accumulation point, at most. Yet Levinson showed that the number
of accumulation points, if superior to one, might be either two or
infinite. The study of the behavior of these graphs in geometrical
terms is also of a certain interest in terms of groups and graph
theory.

\appendix
\addcontentsline{toc}{section}{Enumerations of Cayley graphs}
\section*{Enumerations of Cayley graphs}

In the following sections\footnote{The corresponding graphs and
labeling schemes appear in an additional file which may be found at
\url{http://www.labri.fr/Perso/~renault/research/pages.ps.gz}
($\approx$ 438 ko.).} we enumerate all locally finite planar Cayley
graphs of small inner degree (from 2 to 4). Cayley graphs of degree 2
correspond to cyclic groups. For the other degrees, we enumerate all
possible labeling schemes, each one corresponding to a family of
planar Cayley graphs. For each class, we give the corresponding
presentation and type vector, as long as a representative of this
class. Some classes may have the same representative in terms of
groups or unlabeled graphs, but the corresponding Cayley graphs are
distinct. The following table displays the number of such classes~:

\renewcommand{\arraystretch}{1.2}
$$\begin{array}{|cc||cc|} \hline
\textrm{Degree} & \textrm{Classes} & \textrm{Degree} &
\textrm{Classes} \\ \hline 
1 & 0 & 4 & 26\\ \hline
2 & 1 & 5 & 64\\ \hline
3 & 8 & 6 & 253\\ \hline
\end{array}$$

\end{document}